\documentclass[11pt,epsf]{article}

\usepackage{epsfig}
\usepackage{amssymb}
\usepackage{graphicx}
\usepackage{color}
\usepackage{subfigure}

\begin{document}

\baselineskip 6mm
\renewcommand{\thefootnote}{\fnsymbol{footnote}}


\newcommand{\nc}{\newcommand}
\newcommand{\rnc}{\renewcommand}



\newcommand{\tcb}{\textcolor{blue}}
\newcommand{\tcr}{\textcolor{red}}
\newcommand{\tcg}{\textcolor{green}}


\def\be{\begin{equation}}
\def\ee{\end{equation}}
\def\ba{\begin{array}}
\def\ea{\end{array}}
\def\bea{\begin{eqnarray}}
\def\eea{\end{eqnarray}}
\def\nn{\nonumber\\}


\def\ct{\cite}
\def\la{\label}
\def\eq#1{(\ref{#1})}


\def\a{\alpha}
\def\b{\beta}
\def\g{\gamma}
\def\G{\Gamma}
\def\d{\delta}
\def\D{\Delta}
\def\ep{\epsilon}
\def\e{\eta}
\def\ph{\phi}
\def\Ph{\Phi}
\def\ps{\psi}
\def\Ps{\Psi}
\def\k{\kappa}
\def\l{\lambda}
\def\L{\Lambda}
\def\m{\mu}
\def\n{\nu}
\def\th{\theta}
\def\Th{\Theta}
\def\r{\rho}
\def\s{\sigma}
\def\S{\Sigma}
\def\ta{\tau}
\def\o{\omega}
\def\O{\Omega}
\def\pr{\prime}
\def\fr{\frac}


\def\half{\frac{1}{2}}

\def\goto{\rightarrow}

\def\na{\nabla}
\def\grad{\nabla}
\def\curl{\nabla\times}
\def\div{\nabla\cdot}
\def\pa{\partial}

\def\bra{\left\langle}
\def\ket{\right\rangle}
\def\lb{\left[}
\def\lc{\left\{}
\def\ls{\left(}
\def\lp{\left.}
\def\rp{\right.}
\def\rb{\right]}
\def\rc{\right\}}
\def\rs{\right)}

\def\vac#1{\mid #1 \rangle}


\def\td#1{\tilde{#1}}
\def\check{ \maltese {\bf Check!}}


\def\Tr{{\rm Tr}\,}
\def\det{{\rm det}}


\def\bc#1{\nnindent {\bf $\bullet$ #1} \\ }
\def\ch {$<Check!>$ }
\def\ss {\vspace{1.5cm}}

\begin{titlepage}

\hfill\parbox{5cm} { }

\vspace{25mm}

\begin{center}
{\Large \bf  Correlation functions of the ABJM model}

\vskip 1. cm
  { Bum-Hoon Lee$^{ab}$\footnote{e-mail : bhl@sogang.ac.kr}, 
    Bogeun Gwak$^a$\footnote{e-mail : rasenis@sogang.ac.kr} and
   Chanyong Park$^a$\footnote{e-mail : cyong21@sogang.ac.kr}
  }

\vskip 0.5cm

{\it $^a\,$Center for Quantum Spacetime (CQUeST), Sogang University, Seoul 121-742, Korea}\\
{ \it $^b\,$ Department of Physics,, Sogang University, Seoul 121-742, Korea }

\end{center}

\thispagestyle{empty}

\vskip2cm


\centerline{\bf ABSTRACT} \vskip 4mm

\vspace{1cm}
In the ABJM model, we study the three-point function of two heavy 
operators and an (ir)relevant one. Following the AdS/CFT correspondence, 
the structure constant in the large 't Hooft coupling limit can be factorized into two
parts. One is the structure constant with a marginal operator, which
is fully determined by the physical quantities of heavy operator and gives rise to
the consistent result with the RG analysis. The other can be expressed as the universal 
form depending only on the conformal dimension of an (ir)relevant operator. 
We also investigate the new size effect of a circular string dual to a certain closed spin chain. 
\vspace{2cm}


\end{titlepage}

\renewcommand{\thefootnote}{\arabic{footnote}}
\setcounter{footnote}{0}

\tableofcontents

\section{Introduction}

Applying of the AdS/CFT correspondence to a strongly interacting  
system is one of the active research areas in the theoretical physics. In order to understand
the duality in depth and the gauge theory in the strong coupling regime, we need to clarify
the underlying structure of the AdS/CFT correspondence more clearly.
One of good examples to understand the AdS/CFT correspondence is the 4-dimensional
${\cal N} = 4$ SYM theory dual to the string theory or supergravity in the $AdS_5 \times S^5$
space-time,
in which the conformal symmetry and the integrability play a crucial role to figure out 
the physics of the strongly interacting system \cite{Minahan:2002ve}-\cite{Arutyunov:2006gs}. 
Recently, such works have been generalized to other dimensions.
For example, in order to account for the worldvolume theory of $M$-brane,
the three-dimensional ${\cal N} =8$ Bagger-Lambert-Gustavsson (BLG) model and the Aharony-Bergman-Jafferis-Maldacena 
(ABJM) model for the ${\cal N}=6$ 
Chern-Simons gauge theory have been widely investigated 
\cite{Bagger:2007jr}-\cite{Abbott:2009um}. Moreover, it was shown
that the ABJM model has the dual gravity description in the $AdS_4 \times CP^3$ background
and is integrable at least up to the two-loop level \cite{Krishnan:2008zs}, in the 
$SU(2) \times SU(2)$ subsector even at four-loop level \cite{Minahan:2009aq}. 
In this paper, following \cite{Janik:2010gc}-\cite{Bissi:2011ha}, we will further investigate 
the AdS/CFT correspondence of the ABJM model by calculating the three-point function 
with an (ir)relevant operator.

In the conformal field theory (CFT), if we know the two- and three-point correlation 
functions, the higher point functions can be determined by them. In general, the coordinate 
dependence of two- and three-point functions is unambiguously fixed by the global conformal
symmetry 
\bea
\bra {\cal O}_A (x) {\cal O}_B (y) \ket &=& \frac{\d_{AB}}{|x-y|^{2 \D}} , \nn
\bra {\cal O}_A (x) {\cal O}_B (y) {\cal O}_C (z) \ket &=&  
\frac{a_{ A B C}}{|x-y|^{\D_A + \D_B - \D_C} |x-z|^{\D_A + \D_C - \D_B} 
|y-z|^{\D_B + \D_C - \D_A}} ,
\eea
where $\D_A$ and $a_{ A B C}$ are the conformal dimension and the structure constant respectively.
Actually, since the structure constant is not constrained by the global conformal symmetry,
we should take into account another way. 
Especially, in the strong coupling regime it is almost impossible to fix the structure constant
except in those cases
in which the other symmetries determine the structure constant \cite{Georgiou:2013ff}.
Another exception is the case including a marginal deformation caused by the Lagrangian 
density itself. Since such marginal deformation modifies the coupling constant only, the structure
constant can be determined by a renormalization group (RG) analysis even on the gauge theory side 
\cite{Costa:2010rz}.

In this paper, we will investigate the three-point function with an (ir)relevant operator.
Although the RG analysis is not working anymore, 
the AdS/CFT correspondence can give a clue about the three-point function 
in the large 't Hooft coupling limit.
On the string theory side, the three-point function of two heavy operators and an (ir)relevant
one can be described by a leading interaction between a solitonic string and a massive dilaton 
field propagating on the $AdS_4$ space.  The string theory calculation shows that
the resulting three-point function has the correct coordinate dependence 
expected by the global conformal symmetry and
its structure constant is closely related to that with a marginal operator.
Finally, we suggest a new circular string dual to a closed spin chain. Its two- and three-point
functions show that the size effect of the closed spin chain is suppressed linearly by
$(\bar{J}_1 - J_1 )$,  
while the open spin chain has the finite size effect suppressed exponentially  
\cite{Arutyunov:2006gs,Lee:2008ui,Park:2010vs,Ahn:2011zg}.

The rest of the paper is organized as follows: In Sec. 2, we briefly summarize the results of
the RG analysis with a marginal deformation \cite{Costa:2010rz}. 
In Sec. 3, after evaluating the three-point function of two heavy operators and an (ir)relevant
one, we show that the ratio between structure constants has a universal form independent
of the details of the heavy operator. 
Moreover, we find that the closed spin chain can have the size correction suppressed linearly.
Finally, we finish our work with some concluding remarks.

\section{Marginal deformation of the conformal field theory}

Many authors have recently calculated three-point correlation functions of two heavy 
operators ${\cal O}_H$ and a light one ${\cal O}_L$ in the $AdS_5 \times S^5$ 
background by using the AdS/CFT correspondence 
\cite{Janik:2010gc,Costa:2010rz,Zarembo:2010rr}. For an $N$-point function with only two heavy 
operators, it can be rewritten in a factorized form as a product 
of two- and three-point functions in the large 't Hooft coupling limit \cite{Buchbinder:2010ek}.   
So it is important to know the conformal dimensions of various primary operators
and the structure constants for understanding the CFT in the strong coupling regime. 
However, there is little known about the structure constant except for BPS operators, 
whereas the 
conformal dimensions of heavy operators have widely been studied \cite{Buchbinder:2010vw}. 
The goal of this work is to obtain more insights about the structure constant.

Let us start with summarizing the known results.
Assuming a heavy operator ${\cal O}_H$ with a conformal dimension $\D_H$, its
two-point function is exactly determined by the global conformal symmetry up
to the normalization
\be 
\bra {\cal O}_H (x) {\cal O}_H (y) \ket =  \frac{1}{|x-y|^{2 \D_H}} ,
\ee
where we set the normalization constant $1$. Similar to the two-point function, 
the global conformal symmetry also fixes 
the coordinate dependence of the three-point function with another operator $ {\cal O}_L$
\be
\bra {\cal O}_H (x) {\cal O}_H (y) {\cal O}_L (z) \ket =  
\frac{a_{HHL}}{|x-y|^{2 \D_H - \D_L} |x-z|^{\D_L } 
|y-z|^{\D_L}} ,
\ee
where $\D_L$ denotes the conformal dimension of ${\cal O}_L$. 
Note that since the structure constant $a_{HHL}$ is not constrained by the conformal symmetry,
it should be determined by other methods. If we take account of
a marginal Lagrangian density operator ${\cal O}_{\cal L}$, 
we can exactly decide the structure 
constant through the RG analysis.   
For an Euclidean four-dimensional CFT, the structure constant is associated with the 
conformal dimension of the heavy operator \cite{Costa:2010rz}
\be		\la{rel:ads5}
- g^2 \frac{\pa}{\pa g^2} \D_H = 2 \pi^2 \  a_{HH{\cal L}} ,
\ee
where $g$ denotes a coupling constant and $2 \pi^2$ corresponds to the solid angle
of $S^3$. Since this relation should be satisfied in all coupling regimes,
we can test the AdS/CFT correspondence in the strong coupling limit.  
To do so, we first know what the spectrum corresponding to the operator is.
Following the AdS/CFT correspondence, a heavy operator usually corresponds 
to a solitonic string moving in the dual geometry, whereas a light one is matched with 
a supergravity mode. Especially, a marginal Lagrangian density operator is dual to 
a massless dilaton field.
It was shown by many authors that solitonic strings moving in
the $AdS_5 \times S^5$ background really satisfies the above relation
\cite{Costa:2010rz,Park:2010vs}.

One can easily generalize the relation \eq{rel:ads5} to the $d$-dimensional
CFT case
\be			\la{rel:general}
- g^2 \frac{\pa}{\pa g^2} \D_H = \frac{2 \pi^{d/2}}{\G(d/2)}  \  a_{HH{\cal L}} ,
\ee
where the multiplication factor implies the solid
angle of $S^{d-1}$. In the string theory, there exists another interesting superconformal theory,
the so-called ABJM model, which describes
a three-dimensional ${\cal N}=6$ Chern-Simons theory \cite{Aharony:2008ug}. 
Its dual is the supergravity theory in the $AdS_4 \times CP^3$ background. Since the ABJM model is 
also conformal, one can easily expect that the ABJM model also satisfies \eq{rel:general} as the form
\be		\la{res:structurecon0}
- g^2 \frac{\pa}{\pa g^2} \D_H = 4 \pi \  a_{HH{\cal L}} .
\ee
In \cite{Arnaudov:2010kk}, various solitonic string solutions moving in the $AdS_4 \times CP^3$ background were investigated and it was shown that the RG analysis \eq{res:structurecon0} is really working in the ABJM model as expected. 

\section{Three-point function with an (ir)relevant operator}

In the three-point function of two heavy operators and a marginal one, 
the structure constant can be exactly determined by the RG analysis on the CFT side. 
On the other hand, the same result can also 
be reproduced on the gravity side by evaluating the semiclassical partition function with an 
interaction between a solitonic string and a dual supergravity mode. 
This result is one of the evidences of the AdS/CFT correspondence.
Can we generalize such calculation to the more general cases? 
More specifically, what is the three-point 
function with an (ir)relevant operator instead of a marginal one?
When evaluating the three-point function with an (ir)relevant operator in the strong
coupling regime, the RG analysis and the perturbative calculation are not valid. 
However, the AdS/CFT correspondence can give the answer.
In this section, we will discuss on the three-point functions of various two 
heavy operators and an (ir)relevant one in the large 't Hooft coupling limit.

\subsection{Point-like String in $AdS_4$}

Let us first consider a point particle propagating only on the Euclidean $AdS_4$ space, whose metric in the Poincare patch reads
\be
ds^2 = \fr{ dz^2 + \d_{ij} d x^i d x^j }{z^2}  ,
\ee
where $z$ and $x^i$ correspond to the radial and boundary coordinates respectively. 
The worldline action of a particle is given by the following  
Polyakov-type action
\begin{eqnarray}		\la{act:point}
S_P =\frac{1}{2} \int^{s/2}_{-s/2}d\tau\left(\frac{\dot{x}^i \dot{x}_i
+\dot{z}^2}{z^2}- m^2\right) \, ,
\end{eqnarray}
where the mass of the particle $m$ is very large and $s$ denotes the modular parameter. 
The solution satisfying the equation of motion becomes
\bea		\la{sol:ads4}
x(\tau) &=& R\,\tanh \kappa\tau + x_0 , \nn
z(\tau) &=& \frac{R}{\cosh\kappa\tau} .
\eea
Under the following boundary conditions at an appropriate UV cutoff $\ep$ ($\ep \to 0$)
\be
\lc x(-s/2) ,  z(-s/2) \rc = \lc  0 ,\ep \rc  \quad {\rm and } \quad
\lc x(s/2) ,  z(s/2) \rc = \lc  x_f ,\ep \rc  ,
\ee
the parameters are related to each other
\be		\la{res:bcond}
\kappa\approx\frac{2}{s}\log\frac{x_f}{\epsilon} \quad {\rm and } \quad
x_f \approx 2R\approx 2x_0 ,
\ee
where higher order corrections are ignored.
After regarding the convolution with the relevant wave function \cite{Janik:2010gc,Costa:2010rz}, 
the saddle point $\bar{s}$ is given by
\begin{eqnarray}
\bar{s}=-\frac{2 i}{m}\log\frac{x_f}{\epsilon} .
\end{eqnarray}
At this saddle point, the semiclassical partition function reduces to
\begin{eqnarray}
e^{iS_P}=\left(\frac{\epsilon}{x_f}\right)^{2\Delta_H}
\quad {\rm with} \quad \Delta_H = m\, ,
\end{eqnarray}
where $\D_H$ corresponds to the energy of a massive particle.
Following the AdS/CFT correspondence, $\D_H$ is reinterpreted as the 
conformal dimension of the dual heavy operator and the semiclassical 
partition function is associated with the two-point function
of it. 

In order to evaluate the three-point function with an (ir)relevant operator, we 
first introduce a massive dilaton field propagating on the $AdS_4$ space. If its mass
is denoted by $m_{\ph}$ ($\ll m$), the conformal dimension of the dual light operator
${\cal O}_{\ph}$ is given by
\be
h = \fr{3}{2} + \fr{\sqrt{9 + 4 m_{\ph}^2}}{2} .
\ee
Note that it is allowed for a dilaton field to have a negative mass squared 
in the $AdS$ space, $m_{\ph}^2  \ge - \fr{9}{4}$, where the lower limit corresponds to the 
Breitenlohner-Freedman bound \cite{Breitenlohner:1982bm}.  
The operator dual to a dilaton with a negative or positive mass is relevant or irrelevant 
respectively. A massless dilaton corresponds to the Lagrangian density
operator ${\cal O}_{\cal L}$ studied in the previous section. 
The bulk-boundary propagator of a massive dilaton in the $AdS_4$ 
space is given by \cite{Witten:1998qj,Freedman:1998tz,Balasubramanian:1998de}  
\be
{\cal D}_{{\ph}} \ls z,x ; 0,y \rs = \frac{\G(h)}{\pi^{3/2} \ \G (h - 3/2)} 
\ls \fr{z}{z^2 + (x-y)^2} \rs^h ,
\ee
where a dilaton propagates from the boundary $\lc 0,y\rc$ to the bulk $\lc z,x\rc$.
Then, the three-point function can be expressed by
\be		\la{eq:thrpot}
\bra {\cal O}_{H} (x_f) {\cal O}_{H} (0)  {\cal O}_{\ph}(y)  \ket  = \fr{I}{x_f^{2 \D_H}} ,
\ee
with
\begin{eqnarray}
I &=& \frac{i \ \G(h)}{ 8 \pi^{3/2} \ \G (h - 3/2)}  
\ \int^{\bar{s}/2}_{-\bar{s}/2}d\tau \left(\frac{\dot{x}^i \dot{x}_i+\dot{z}^2}{z^2}-m^2\right)\left(\frac{z}{z^2+(x-y)^2}\right)^h \nn
&=& -
\frac{m}{2^{h +2} \pi}  \frac{ \G \ls \frac{h}{2} \rs \G \ls h \rs}{\G \ls \frac{h +1}{2} \rs 
\G \ls h - \frac{3}{2} \rs} 
\ \frac{ 1}{x_f^{-{h}} \ | x_f - y | ^{h} \ y^{h}} + \cdots,
\end{eqnarray}
where the solutions in \eq{sol:ads4} are used and the ellipsis implies higher order corrections
in the large $\bar{s}$ limit.
In \eq{eq:thrpot}, $I$ implies the interaction between a solitonic string and a massive 
dilaton field.

For $m_{\ph}=0$, the light operator is marginal and the three-point function simply reduces to
\be
\bra {\cal O}_{H} (x_f) {\cal O}_{H} (0)  {\cal O}_{\cal L} (y)  \ket  = -\frac{ m}{16 \pi} \frac{1}{ x_f^{2 \D_H- 3} |x_f-y|^3 y^3} ,
\ee
which coincides with the result in \cite{Arnaudov:2010kk}.
Assuming that $\D_H = m \sim \sqrt{g}$ \cite{Costa:2010rz}, we can easily check that the structure constant satisfies the result of the RG analysis \eq{res:structurecon0}
\be
- g^2 \frac{\pa \D_H}{\pa g^2} = - \frac{m}{4} = 4 \pi a_{HH{\cal L}}  .
\ee
For $m_{\ph} \ne 0$, the three-point function can be summarized  to
\be
\bra {\cal O}_{H} (x_f) {\cal O}_{H} (0)  {\cal O}_{\ph}(y)  \ket = 
\ \frac{ a_{HH{\ph}} }{x_f^{2 \D_H-{h}} \ | x_f - y | ^{h} \ y^{h}}   ,
\ee
with 
\be
a_{HH{\ph}} = - \frac{m}{2^{h +2} \pi}  \frac{ \G \ls \frac{h}{2} \rs \G \ls h \rs}{\G \ls \frac{h +1}{2} \rs 
\G \ls h - \frac{3}{2} \rs}   ,
\ee
which shows the coordinate dependence expected by the global conformal symmetry.

\subsection{A circular string wrapped in $\th$ }

Now, consider a solitonic string moving in the $AdS_4 \times S^3$ background
which is a subspace of the $AdS_4 \times CP^3$ geometry dual to the ABJM model. 
Here, $S^3$ represents the diagonal subspace of $CP^3$ \cite{Lee:2008ui,Abbott:2009um}
\begin{eqnarray} 		\la{met:s3}
ds^2=\frac{1}{4} \ls d\theta^2 + \sin^2\theta d\phi_1^2+ \cos^2\theta d\phi_2^2 \rs .
\end{eqnarray}
Under the ansatz for a circular string extended 
in $\th$ with rotations in $\ph_1$ and $\ph_2$,
\begin{eqnarray}
\theta = \sigma\,,\,\,\phi_1=\omega_1\tau\,,\,\,\phi_2=\omega_2\tau\, ,
\end{eqnarray}
the Polyakov string action becomes
\be		\la{act:gen}
S = \frac{T}{2} \int^{s/2}_{-s/2}  d\tau \int_0^{2 \pi} d \s \lb \frac{\dot{x}^i \dot{x}_i
+\dot{z}^2}{z^2}  - \th'^2 + \sin^2 \th \ \dot{\ph}_1^2 + \cos^2 \th \ \dot{\ph}_2^2 \rb  ,
\ee
where the dot and the prime represent the derivatives with respect to $\ta$ and $\s$ respectively.
In \eq{act:gen}, the first two terms describe the motion of the string in the $AdS_4$ space.
Since all solitonic strings studied in the paper behave like a point particle 
in the $AdS_4$ space, their
solutions are also given by \eq{sol:ads4}.
Note that the string tension $T$ in the $AdS_4 \times CP^3$ space 
is associated with the 't Hooft coupling constant $\l$
\cite{Lee:2008ui}
\be
T = \sqrt{\frac{\l}{2}}  = 2 g  ,
\ee 
where $g$ is the coupling constant appearing in Sec. 2.
Following \cite{Janik:2010gc,Costa:2010rz}, 
at the saddle point 
\be
\bar{s} = \frac{ 2 \sqrt{2}}{i \sqrt{2 + \o_1^2 + \o_2^2}}   \log \ls \frac{x_f}{\ep} \rs ,
\ee
the semiclassical partition function becomes
\be \la{res:cirtwo}
e^{i S} =  \ls \frac{\ep}{x_f} \rs^{2 \D_H} ,
\ee
where the energy of a circular string $\D_H$ reads
\be
\D_H = \sqrt{2} \ \sqrt{ J_1^2 +  J_2^2 + 2 \pi^2 T^2 }  ,
\ee
in terms of the angular momenta
\be
J_1 = \pi T \o_1 \quad {\rm and } \quad J_2 = \pi T \o_2 .
\ee
If a dilaton field is massless, the RG analysis \eq{res:structurecon0} expects 
the structure constant to be
\be		\la{res:scRG}
4 \pi \  a_{AA{\cal L}}  
= - \frac{\sqrt{2} \pi^2 T^2}{\sqrt{J_1^2 +  J_2^2 + 2 \pi^2 T^2}} .
\ee

On the string theory side, the result \eq{res:scRG} can also be reproduced from the three-point function with a general light operator. The general form of the three-point function with
an (ir)relevant operator is given by
\be \la{eq:threepointfn}
\bra {\cal O}_{H} (0) {\cal O}_{H} (x_f) {\cal O}_{\ph} (y) \ket \ =\
\frac{I}{x_f^{2 \D_H}} ,
\ee
with
\begin{eqnarray}			\la{int:cirth}
I &=& \frac{i \ \G(h)}{ 4 \pi^{3/2} \ \G (h - 3/2)} \nn  
&& \times \int^{\bar{s}/2}_{-\bar{s}/2}  d\tau \int_0^{2 \pi} d \s \ls \frac{\dot{x}^i 
\dot{x}_i
+\dot{z}^2}{z^2}  - \th'^2 + \sin^2 \th \ \dot{\ph}_1^2 + \cos^2 \th \ \dot{\ph}_2^2 \rs 
\left(\frac{z}{z^2+(x-y)^2}\right)^h  \la{eq:threeptfn} .
\end{eqnarray}
After integrating \eq{int:cirth}, the leading term of the three-point function
gives rise to
\be
\bra {\cal O}_{H} (0) {\cal O}_{H} (x_f) {\cal O}_{\ph} (y)\ket  = -
\frac{\pi T^2}{2^{h - 1/2} \sqrt{J_1^2 + J_2^2 + 2 \pi^2 T^2}}   
\frac{ \G \ls \frac{h}{2} \rs \G \ls h \rs}{\G \ls \frac{h +1}{2} \rs  \G \ls h - \frac{3}{2} \rs} 
\ \frac{ 1}{x_f^{2 \D_H-{h}} \ | x_f - y | ^{h} \ y^{h}} .
\ee
Its coordinate dependence is the exact form expected by the CFT. 
For $m_{\ph}=0$, the dual light operator becomes marginal $h=3$ and the three-point function reduces 
to
\be
\bra {\cal O}_{H} (0) {\cal O}_{H} (x_f) {\cal O}_{\cal L} (y) \ket   =
- \frac{\sqrt{2} \pi T^2}{4 \sqrt{ J_1^2+ J_2^2 + 2 \pi^2 T^2 }}  \ \fr{1}{x_f^{2 \D_H-3} | x_f - y | ^3 y^3} ,
\ee
where we can see that the structure constant perfectly coincides with \eq{res:scRG} derived by the RG analysis.
Comparing above two structure constants yields the following ratio 
\be			\la{res:uninversal}
\fr{a_{HH{\ph}}}{a_{HH{\cal L}}}
= \frac{1}{2^{h-2}}  
\frac{ \G \ls \frac{h}{2} \rs \G \ls h \rs}{\G \ls \frac{h +1}{2} \rs  \G \ls h 
- \frac{3}{2} \rs}  ,
\ee
in which the result shows a universal form in that it does not contain 
any information about the heavy operator.
This implies that in the large 't Hooft coupling limit the structure constant with
an (ir)relevant operator can be factorized into two parts. One is $a_{HH{\cal L}}$
determined by the details of heavy operator and the other depends on the conformal
dimension of the light operator only. 
In the following sections, we will check
this new feature of the structure constant with more complicated solitonic strings.

\subsection{Dyonic Magnon}

In this section, we will take into account a more nontrivial solitonic string called a dyonic magnon. 
In the $AdS_5 \times S^5$ background dual to the $N=4$ Super Yang-Mills theory, 
the three-point function of two dyonic magnons and a marginal operator has been 
investigated \cite{Costa:2010rz,Buchbinder:2010vw,Arnaudov:2010kk,Park:2010vs,Ahn:2011zg}. 
This work was also generalized to the ABJM model \cite{Park:2010vs}.
Here, we will further study the three-point function with an (ir)relevant 
operator and check the universal behavior of the structure constant ratio.

A dyonic magnon corresponds to a bound state of magnons in the spin chain model,
which can be described on the string theory side by a solitonic string rotating on $S^3 \subset CP^3$.
The ansatz for a dyonic string is given by
\be
\th = \th(y) \quad , \quad \ph_1 = \n_1 \ta + g_1 (y) \quad {\rm and } \quad 
\ph_2 = \n_2 \ta + g_2 (y) ,  
\ee
with 
\be
y = a \ta + b \s .
\ee
The rotational symmetries in $\ph_1$ and $\ph_2$ give rise to
\bea
g_1' &=& \fr{1}{b^2 - a^2} \ls a \n_1 - \fr{c_1}{\sin^2 \th} \rs , \nn
g_2' &=& \fr{1}{b^2 - a^2} \ls a \n_2 - \fr{c_2}{\cos^2 \th} \rs ,
\eea
where $c_1$ and $c_2$ are integration constants and the prime means the derivative 
with respect to $y$. Here, we take $b^2 > a^2$ and $c_2 =0$ to obtain a dyonic magnon
solution \cite{Lee:2008ui,Park:2010vs}. 
Using the Virasoro constraints, the equation
of motion for $\th$ can be rewritten as the first order differential equation form
\be	\la{eq:thmm}
\theta'^2 = \frac{b^2(\nu_1^2-\nu_2^2)}{(b^2-a^2)^2\sin^2\theta}\left(\sin^2\theta_{max}
-\sin^2\theta\right)\left( \sin^2\theta - \sin^2\theta_{min}\right),		
\ee
with
\bea
\sin^2 \th_{max}  &=& \frac{c_1}{a \n_1}  ,  \la{res:thmax} \\
\sin^2 \th_{min} &=&  \frac{a \n_1 c_1}{b^2(\nu_1^2-\nu_2^2)} .
\eea
From now on, we concentrate on the infinite size limit ($J_1 \to \infty$), which can be accomplished
by setting $c_1 = a \n_1$ ($\sin \th_{max} = 1$).
After the convolution, the semiclassical partition function is represented as
\begin{eqnarray}
e^{iS}=\left(\frac{\epsilon}{x_f}\right)^{2 \D_H} ,
\end{eqnarray}
with 
\be   	\la{res:magdim}
\D_H =  J_1 + \sqrt{J_2^2 + 4T^2 \sin^2 \frac{p}{2}} ,
\ee
which was evaluated at the saddle point
\be
\bar{s}=- \frac{2 i}{\n_1}\log\frac{x_f}{\epsilon} .
\ee
In \eq{res:magdim}, $J_i$ ($i=1,2$) means the angular momentum in the $\ph_i$-direction
and $p$ is the worldsheet momentum which can be reinterpreted
in the target space as the angle difference of two ends of a string
\cite{Lee:2008ui}.

Following the method used in the previous sections, one can finally find
the following three-point function of two dyonic magnons and an (ir)relevant operator
\be
\bra {\cal O}_{H} (0) {\cal O}_{H} (x_f) {\cal O}_{\ph} (y) \ket  =
 \frac{a_{HH{\ph}}}{x_f^{2 \D_H -{h}} \ | x_f - y | ^{h} \ y^{h}} .
\ee
where the structure constant is given by
\be
a_{HH{\ph}} =  -
\fr{T^2 \sin^2 (p/2)}{2^{h-1} \pi \sqrt{J_2^2 + 4 T^2 \sin^2 (p/2)}}
\frac{ \G \ls \frac{h}{2} \rs \G \ls h \rs}{\G \ls \frac{h +1}{2} \rs \G \ls h - \frac{3}{2} \rs} .
\ee
For $m_{\ph}=0$, it reduces to
\be
a_{HH{\cal L}} 
= - \fr{T^2 \sin^2 (p/2)}{2 \pi \sqrt{J_2^2 + 4 T^2 \sin^2 (p/2)}}  ,
\ee 
and satisfies the RG analysis \eq{res:structurecon0}.
Furthermore, the ratio of the above structure constants shows the same universal
form in \eq{res:uninversal}.

\subsection{A circular string wrapped in $\ph_1$}

The motivation of this section is not only to check the universality mentioned before  but also
to investigate a new size effect of the closed spin chain. Usually, a circular string corresponds
to a closed spin chain in the dual CFT whereas a magnon is dual to a open spin
chain. If the magnon's size $J_1$ in \eq{res:magdim} is large but 
finite, there is an additional finite size effect on the conformal dimension which is 
exponentially suppressed like $e^{- J_1}$ \cite{Park:2010vs,Ahn:2011zg}. 

What is the size effect of the closed spin chain? 
In order to answer this question, we think of another circular string which is wrapped in $\ph_1$ and rotating in $\ph_1$ and $\ph_2$. Then, 
the appropriate ansatz is given by
\be
\ph_1 = \o_1 \ta + w \s \quad , \quad \ph_2 = \o_2 \ta \quad {\rm and} \quad \th = 
\th_0 ,
\ee
where $w$ is the winding number and $0 \le \s < 2 \pi$. 
We assume that the position of the string in $\th$ is fixed to $\th_0$ and that 
two angular velocities $\o_1$ and $\o_2$ are finite.
For $\th_0 = \pi/2$, the above ansatz reduces to one wrapping the 
equator of $S^2$. From the string action
\bea	\la{act:general}
S &=& \frac{T}{2} \int^{s/2}_{-s/2}  d\tau \int_0^{2 \pi} d \s \ \lb \ \ \fr{ \ls \dot{x^i} \rs^2
- \ls {x^i}' \rs^2 +\dot{z}^2 - {z'}^2 }{z^2}  \rp \nn
&& \lp  \qquad \qquad \qquad \qquad \qquad + \ \dot{\th}^2 - \th'^2 
+ \sin^2 \th \ls \dot{\ph}_1^2  - {\ph'}_1^2 \rs + \cos^2 \th \ls  \dot{\ph}_2^2
 - {\ph'}_2^2 \rs  \ \rb  ,
\eea
after regarding the convolution contribution and setting $\sin \th_0 = \sqrt{1 - \d^2}$, 
we find a saddle point at
\be
\bar{s} = \frac{2}{ i \sqrt{ (w^2 + \o_1^2) (1 - \d^2) + \o_2^2 \ \d^2 } }
 \log \frac{ x_f}{ \ep }.
\ee
At this point, the semiclassical partition function leads to
\be
e^{i S} =  \ls \frac{\ep}{x_f} \rs^{2 \D_H } ,
\ee
with
\bea			\la{res:condimcirph}
\D_H 
&=&   2 \pi T \sqrt{   \frac{J_1^2}{4 \pi^2 T^2 (1 - \d^2)} + (1 - \d^2) w^2 
+ \frac{J_2^2}{4 \pi^2 T^2 \d^2} } ,
\eea
where two angular momenta are defined by
\bea
J_1 &=& 2 \pi T \o_1 (1 - \d^2)  , \nn
J_2 &=& 2 \pi T \o_2  \d^2  .
\eea
For understanding the above result in more depth, we first take account of the 
case $\d = 0$. In this case, $S^3$ reduces to $S^2$ and the conformal dimension of
a circular string is given by
\be
\bar{\D}_H = \sqrt{  \bar{J}_1^2 + 4 \pi^2 T^2 w^2 } ,
\ee
where the bar symbol means a quantity defined on $S^2$, like $\bar{J}_1 \equiv 2 \pi T \o_1$.  

In the large 't Hooft coupling limit ($T \gg 1$), $\D_H$ and $J_1$ are large ($\sim T$) but $J_2$ 
is proportional to $T \d^2$. If we define $\bar{J}_2 \equiv 2 \pi T \o_2$, 
the conformal dimension in \eq{res:condimcirph}
can be expanded near the equator of $S^2$ ($\d \ll 1$) to
\be			\la{res:circonf}
\D_H =   \sqrt{  \bar{J}_1^2 + 4 \pi^2 T^2 w^2 } +
\frac{\bar{J}_2^2 -  \bar{J}_1^2 - 4 \pi^2 T^2 w^2}{2  \sqrt{  \bar{J}_1^2 + 4 \pi^2 T^2 w^2 } } 
\   \d^2 + \cdots .
\ee
Here, the first term is nothing but the conformal dimension of the circular string
living on the equator of $S^2$ and the second is the leading size effect 
caused by the change of the string length. Near the equator of $S^2$, the string length $l$ 
can be expanded to
\be
l = 2 \pi \sqrt{1 - \d^2} \approx 2 \pi - \pi \d^2 ,
\ee
where $\pi \d^2$ parameterizes the deviation of its string length from the one of a string wrapping around the equator.
Rewriting $\d^2$ in terms of $J_1$ and $\bar{J}_1$
\be
\d^2 = \frac{\bar{J}_1 - J_1}{\bar{J}_1} ,
\ee
the conformal dimension of the circular string becomes
\be
\D_H  =  \sqrt{ \bar{J}_1^2 + 4 \pi^2 T^2 w^2 
+ \ls \bar{J}_2^2 -  \bar{J}_1^2 - 4 \pi^2 T^2 w^2 \rs   \frac{\bar{J}_1 - J_1 }{\bar{J}_1}} .
\ee
In the dual closed spin chain picture, this shows that 
the leading size effect on the conformal dimension is linearly proportional to $\bar{J}_1 - J_1$
unlike the magnon case.

After some calculations, the three-point function with an (ir)relevant operator
finally becomes
\bea
\bra {\cal O}_{H} (0) {\cal O}_{H} (x_f) {\cal O}_{\ph} (y) \ket   &=&
- \frac{ \pi  T^2 \ w^2  \  J_1 }{2^{h-1}   \bar{J}_1
\sqrt{\bar{J}_1^2 + 4 \pi^2 T^2 w^2 
+ \ls \bar{J}_2^2 -  \bar{J}_1^2 - 4 \pi^2 T^2 w^2 \rs  \frac{\bar{J}_1 - J_1 }{\bar{J}_1}} }
\nn
&&  \times \ 
\frac{ \G \ls \frac{h}{2} \rs \G \ls h \rs}{\G \ls \frac{h +1}{2} \rs \G \ls h - \frac{3}{2} \rs} 
\ \frac{ 1}{x_f^{2 \D-{h}}  | x_f - y | ^{h}  y^{h}} .
\eea 
For $m_{\ph}=0$, it simply reduces to the three-point function with a marginal operator 
with the following structure constant  
\be
a_{HH{\cal L}} = - \frac{ \pi  T^2 \ w^2  \ J_1}{2  \bar{J}_1
\sqrt{\bar{J}_1^2 + 4 \pi^2 T^2 w^2 
+ \ls \bar{J}_2^2 -  \bar{J}_1^2 - 4 \pi^2 T^2 w^2 \rs \frac{\bar{J}_1 - J_1 }{\bar{J}_1}} } ,
\ee
which also satisfies the RG analysis \eq{res:structurecon0}. Furthermore, the universality
in \eq{res:uninversal} is still preserved.
For a small $\d$, the structure constant is also expanded to
\bea
a_{HH{\cal L}} \approx 
 - \frac{\pi T^2 w^2}{2 \sqrt{\bar{J}_1^2 + 4 \pi^2 T^2 w^2} } 
+ \frac{ \pi T^2 w^2 (\bar{J}_1^2 + \bar{J}_2^2 + 4 \pi^2 T^2 w^2) }{4 \ls \bar{J}_1^2 + 4 \pi^2 T^2 w^2 \rs^{3/2}} \  \frac{\bar{J}_1 - J_1 }{\bar{J}_1}  ,
\eea 
in which the first is the structure constant of a circular string living on the equator of $S^2$ and the second is the leading size effect caused by the change of the string length respectively.

\section{Conclusions}

In the strong coupling regime, 
it is almost impossible to calculate a general three-point function by using the traditional way of
the
QFT. Anyway, there are several exceptions. If there is an additional symmetry, 
the three-point function with its current operator can be determined 
by the Ward identity even in the strong coupling regime \cite{Georgiou:2013ff}. 
Another possible example is the
three-point function with a marginal operator, more specifically a Lagrangian density operator.
On the CFT side, the three-point function of two heavy operators and a marginal one
can be evaluated by the RG analysis, which shows that the structure constant $a_{HH{\cal L}}$
is related to
the derivative of the conformal dimension of the heavy operator. Following the AdS/CFT  
correspondence, we investigated the three-point function of the ABJM model 
in the strong coupling regime and showed that the string calculation really 
leads to the consistent result with the RG analysis.

In this paper, we further investigated the three-point function with an (ir)relevant operator 
\be
\bra {\cal O}_{H} (0) {\cal O}_{H} (x_f) {\cal O}_{\ph} \ket  =
 \frac{ a_{HH{\ph}} }{x_f^{2 \D-{h}} \ | x_f - y | ^{h} \ y^{h}} ,
\ee
where the structure constant is given by
\be   		
a_{HH{\ph}}
= \frac{1}{2^{h-2}}  
\frac{ \G \ls \frac{h}{2} \rs \G \ls h \rs}{\G \ls \frac{h +1}{2} \rs  \G \ls h - \frac{3}{2} \rs} 
a_{HH{\cal L}} .
\ee
This result shows that the coordinate dependence is the form exactly expected by the conformal symmetry. 
Interestingly, the above structure constant is closely related to that with a marginal operator
and their ratio has a universal feature not depending on the details of the heavy operator. 
We have checked this universality with the various heavy operators corresponding to the solitonic
strings moving in the $AdS_4 \times CP^3$ space.   
These results can easily be generalized to the higher dimensional cases like the
$AdS_5 \times S^5$ background, in which the definition of the string tension should be modified \cite{Lee:2008ui}.
Finally, we found a solitonic string dual to a certain closed spin chain in the dual CFT
and studied the new size effect of it. In the large 't Hooft coupling limit, the size effect of the closed spin chain is suppressed linearly by $ (\bar{J}_1 - J_1 )$,
while the open spin chain described by a magnon has the exponential suppression
$ e^{- J_1}$.

\vspace{1cm}

{\bf Acknowledgement}

This work was supported by the National Research Foundation of Korea(NRF) grant funded by
the Korea government(MEST) through the Center for Quantum Spacetime(CQUeST) of Sogang
University with grant number 2005-0049409. C. Park was also
supported by Basic Science Research Program through the
National Research Foundation of Korea(NRF) funded by the Ministry of
Education, Science and Technology(2010-0022369).
This research was also supported by the International Research and Development Program of the National Research Foundation of Korea (NRF) funded by the Ministry of Education, Science and Technology(MEST) of Korea(Grant number: 670552).

\end{document}